# Detecting Network Anomalies using Rule-based machine learning within SNMP-MIB dataset

Abdalrahman Hwoij[*1], Mouhammd Al-kasassbeh[2], Mustafa Al-Fayoumi[3]

Department of Computer Science, Princess Sumaya University for Technology
Amman, Jordan

[*1]Abd20178080@std.psut.edu.jo; [2] m.alkasassbeh@psut.edu.jo; [3] m.AlFayoumi@psut.edu.jo

*Abstract-* **One of the most effective threats that targeting cybercriminals to limit network performance is Denial of Service (DOS) attack. Thus, data security, completeness and efficiency could be greatly damaged by this type of attacks. This paper developed a network traffic system that relies on adopted dataset to differentiate the DOS attacks from normal traffic. The detection model is built with five Rule-based machine learning classifiers (DecisionTable, JRip, OneR, PART and ZeroR). The findings have shown that the ICMP variables are implemented in the identification of ICMP attack, HTTP flood attack, and Slowloris at a high accuracy of approximately 99.7% using PART classifier. In addition, PART classifier has succeeded in classifying normal traffic from different DOS attacks at 100%.**

## I. INTRODUCTION

Recently, the security risks to the network and to the stability of the Information Technology (IT) business sector have gradually been growing. Denial of Service (DoS) attack is one of the most powerful and effective attacks. Many DoS attacks against various organizations have been carried out worldwide since summer 1999. Such assaults crippled the economy and forced several firms to exit the sector. Online banking services for nine big banks in USA have been accessible since September 2012. Banks have suffered a variety of these attacks [1], contributing to financial losses, and the recovery of operations and a return to normal activity is difficult and costly. Consequently, it is important to research and combat these assaults. Intrusion detection system (IDS) is a process of network traffic monitoring and comparing with normal operations to detect signs of intrusion from any abnormal behavior or action. Intrusion is an operation that is harmful in order to jeopardize the privacy, credibility and functionality of network components with a view to breaching network security policy [2]. The two main detection methods in IDS are Misuse Intrusion Detection (MID) and Anomaly Intrusion Detection (AID). Signature-based detection is also known as MID. It is used to track threats, but it is likely to miss and not identify new types of attacks. AID utilizes abnormal behavior on servers or networks to spot abnormalities. This needs to build typical habits that are accumulated during normal operations over a period of time. The AID devices are useful for the identification of unusual threats without prior knowledge. Researchers are currently focused primarily on AID, as known and unknown attacks can be detected. An overview of IDS systems is presented in Figure 1.

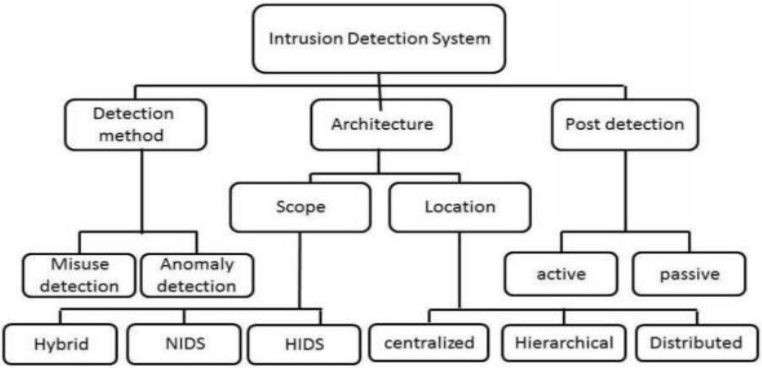

*Figure 1. IDS systems* [3]

This research focuses on methods for network anomaly identification. In the case of any attacker on a network or due to network congestion, network anomalies are a disturbance in network behavior. The ordinary operation for network services is compromised by these anomalous activities. Different factors such as the size of network data to calculate, network traffic volume, and the number of applications working on a network will character typical network behavior [3]. Traffic characteristics defining network traffic actions are used to identify network trafficable irregularities and establish a standard template of network traffic. This model can be generated from the training data using models or mathematical models. A Network Attack is any mechanism or tool used to try to breach network security against computer networks [4]. Different steps in the assaults, beginning with the initial motivation of the attackers to the final performance, are called by attackers [5]. The most appealing form of attack on attackers is a DOS strike. This research focuses on the prevention of threats and floods through DOS and brute force assaults. Transmission Control Protocol-Synchronize (TCP-SYN) attack, User Datagram Protocol (UDP) Flood an gripping, ICMP-Echo Attack of Internet Control Protocol, Flood Attack, Slowloris Attack and Slowpost Attack are the most common types of DOS flooding attacks. On the other hand, attacker attempt to log into a user account with brute force (password crack). The attackers try to constantly attempt to use various passwords on the victim to get the correct password. user passwords, such as "admin", "1234567890", etc. were very weak, through brute force attack, the right password will be quickly revealed. Through brute force attack, the correct code will quickly be divulged. The full data collection and attacks used by myself and other authors in this analysis have been collected from previous research [6]. Network abnormalities [2] [7] are recognized as various methods, such as statistical, rule-modeling, threshold and machine learning (ML). Statistically speaking, numerical anomalies are detected; based on data analysis, it is determined whether the behavior is normal or unusual. Anomalous behavior is observed when system-defined rules are ignored in the rule modeling method. The irregular behavior is observed in the threshold method when limits are exceeded for information variability tracking. Using past data or actions, the latter method builds a prototype. In normal and abnormal circumstances, common and uncommon data is collected from the network. Data are then marked for model training. The main advantage of the ML method is that it is suitable for unknown attacks and that it can be adapted to any network traffic changes. ML is the most common way of detecting abnormality [8].

The contributions in this paper include:

1) Investigating the Intrusion Detection (ID) problem using five Rule-Based machine learning algorithms namely, Decision Table, JRIP, oneR, PART and zeroR.

2) Using these algorithms on a modern real Management Information Based (MIB) dataset generated by Al-Kasassbeh et al [6]. which is obtained from real-life conditions.

3) Improve the accuracy of the detection system using a variety of choice techniques like Precision, Recall, F-Measure and Accuracy.

4) The experiments show that the five methods of selection improved the performance, especially accuracy of classification.

Distributed Denial-of-Service attack (DDoS) is a deliberate attempt to disrupt normal traffic by an Internet traffic jamming the target or its associated networks. By using multiple compromised computer systems as source of attack traffic DDoS attacks will be effective. Computers and other networked assets, including Internet of Things (IoT) phones, can be part of managed machines. From a high standard, a DDoS assault is like a bottleneck that is created by the bridge and blocks normal traffic at its desired location. An attack by DDoS requires an attacker to control a network of online machines for an attack. The malware becomes compromised by computers and other systems (such as IoT devices), each one of them transformed into a bot (or zombie). The hacker is then remotely controlled by the bots group called a botnet. When a botnet is established, the attacker can direct the machines through a method of remote control by sending updated instructions to each bot. That bot can respond through transmission of requests to the aim, potentially resulting in a Denial-of-Service DoS of normal traffic, to the IP address of the victims aimed at the botnet. Because every bot is a legitimate Internet device, it can be difficult to separate attack traffic from normal traffic [9].

Types of DDoS attacks:

a) TCP-SYN Flood:

In a TCP-SYN flood attack, the assailant sends TCP-SYN repeated packets, often using a fake IP address, to any port on the target Server. Without knowing about the attack, the server receives several, apparently legitimate, communication requests. The SYN-ACK packet answers every attempt from every open port. Either the malicious user does not submit the ACK anticipated, or –if the IP is spoofed–it does not first receive the SYN-ACK. In any event, the assaulting server is waiting for some time to identify the SYN-ACK file [9].

b) UDP Flood:

In UDP Flood DDoS attackers use a large source IP range to send high-speed UDP (User Datagram Protocol) packets at a very high packet rate. Many incoming UDP packets overwhelmed the victim network. This assault typically absorbs network resources and capacity, which overload the network to the extent where it falls offline [9].

c) ICMP-ECHO:

The ICMP flood attack is also a ping flood attack when an attacker tries to overwhelm a targeted device with the ICMP echo requests (pings). an ICMP flood attack is used for the attacker. In order to diagnose the sanitary, connectivity and connections between the sender and the device, ICMP echo-request and reply messages are normally utilized as a network device to pin. The network is forced to answer with the same amount of reaction packets by inundating the target with query packets. It renders the goal impractical to ordinary traffic [9].

d) HTTP Flood:

HTTP Flood attack exploits seemingly-legitimate HTTP GET or POST requests to attack a web server or application. HTTP flood attacks are volumetric attacks, often using a botnet "zombie army" a group of Internet-connected computers, each of which has been maliciously taken over, usually with the assistance of malware like Trojan Horses. A sophisticated Layer 7 attack, HTTP floods do not use malformed packets, spoofing or reflection techniques, and require less bandwidth than other attacks to bring down the targeted site or server. As such, they demand more in-depth understanding about the targeted site or application, and each attack must be specially-crafted to be effective. This makes HTTP flood attacks significantly harder to detect and block [9].

e) Slowloris Attack:

This attack is also known as an attack by the Slow Header. The attacker sends sessions with a high workload request with an unexpected IP address. These requests are partial requests from the HTTP header which are slow to update, grow quickly, continuously and never close. The attack continues until the requests take up all available sockets and the web server cannot have a legitimate connection. The attack will lead to a crash on a web server using only a few computers, without any side effect on other systems and ports. The attack will trigger an assault [10].

f) Slowpost Attack:

This attack is referred to as a Slow Request Bodies and came in 2010 for the first time. This assault is close to a Slowloris attack by submitting extremely workload requests to access web servers by the attackers. The assailant will send a full HTTP header request for this attack, which will define the contents in the post message field as the request is sent for normal traffic. The information is then sent every two minutes to fill in a message box at a frequency of 1 byte and the database then waits until each message body is finished and web services are rejected [10].

An Intrusion Detection System (IDS) is a system that monitors suspicious activity traffic and issues alertness when such activity is discovered. It is a software application that scans a network or system for damaging activities or breaches of policies. While intrusion detection systems screen networks for potential malicious activity, they also are prepared to provide false alarms. Organizations therefore need to optimize their IDS services when they are implemented first. That ensures that intrusion detection systems are properly established to know how regular network traffic looks compared to malevolent behavior. IDS is classified into two types, Network Intrusion Detection System (NIDS) and Host Intrusion Detection System (HIDS). Network Intrusion Detection Systems (NIDS) are designed to traffic from all network devices at a planned point within the network. This tracks the traffic flow through the whole subnet which correlates with the information that is transmitted on the subnet to capture identified threats. After identifying an attack or observing unusual behavior, the alarm can be forwarded to the administrator, Host intrusion detection systems (HIDS) operate on different servers or network devices. A HIDS monitors only incoming and outgoing device packets and alerts the manager to suspicious activity or malice detection. This takes a snapshot and compares existing system files to the previous snapshot. Prevention of intrusion is a proactive approach to network security to detect and respond quickly to potential threats. Like an IDS system, the IPS monitors network traffic. Network traffic controls the traffic. Nonetheless, since a vulnerability can be created very quickly following the attacker's entry, intrusion prevention programs may take immediate action based on a set of network administrator guidelines. Intrusion prevention system (IPS) is a preventive approach to network security to identify and respond quickly to potential threats. Like an IDS system, the IPS monitors network traffic. Network traffic controls the traffic. Nonetheless, since a vulnerability can be created very quickly following the attacker's entry, intrusion prevention programs may take immediate action based on a set of network administrator guidelines.

Proper network management is a key component of effective network management. Network administrators need tools to monitor the functionality, linkages and services of the network devices. In the field of network management solutions SNMP is a popular protocol to use a TCP / IP protocol suite to manage network devices on the Internet. SNMP is a User Datagram Protocol (UDP) system layer protocol. It's used to customize and gather network interface data including Standard PCs, switches, servers and routers [11]. Since it was introduced in the 1980s, SNMP have, because of its flexibility, been the method to handling network equipment across different network styles. Most current intrusion detection research's depends on the analysis of raw packet information to assess computer systems and network security status, resulting in a significant processing burden and late detection time [12]. Additional network traffic data sources supplied by multiple network administration protocols such as CIMP, remote network surveillance (RMON) and the protocol that is the most widely used SNMP.

As described above, the SNMP supports variables that are equal to the system level traffic information. Such Network System Data can be tracked passively and used to classify network activity and therefore for monitoring of network abnormalities [11]. There are three versions of SNMP protocol, the first version SNMPv1 found in RFC 1157, the second version was introduced with RFC 1441and the latest version described in RFCs 3410.

An Information Base of Management (MIB) is a database for objects to be handled in the network system. MIB is frequently accompanied by SNMP. Since SNMP has already been developed and introduced for all network devices, we have therefore chosen the SNMP for the monitoring of network attacks as our protocol to provide MIB statistic data easily obtained for study. The use of the fine grain information from MIB to identify irregularities will eliminate certain of the problems of network intrusion detection. A unique MIB factor is somehow influenced for each malicious activity that arises. Thus, SNMP-MIB information is a type of anomaly detection indicators. For the more accurate detection of network abnormalities, proper SNMP-MIB variables must be chosen as there is no single variable that captures all network abnormalities, minimizing the number of MIB variables involved and maximizing the anomaly range covered [12].

This paper is organized as follows. The introduction is presented in Section 1. Section 2 provides a literature review that includes different techniques for solving the IDS problem. Section 3 Briefly outlines the essence of the research data set with the five ML algorithms namely, Decision Table, JRIP, oneR, PART and zeroR. Section 4 describes the experiments of evaluating the model developed. Experimental results are shown in Section 5. Finally, the conclusion are shown in Section 6.

## II. RELATED WORK

During the past decade, several scientists have examined significant phenomena and assaults on computer networks. Anomaly detection surveys were presented and many different approaches were proposed and implemented, In the field of detection of network abnormalities, many studies have used SNMP-MIB data. Some researchers have introduced methods based on statistical analysis of MIB data, while others have used machine learning techniques for network attacks and other anomalies recently.

In [13]. The authors analyse dataset in order to detect application-based DoS attacks using Random Forest, Extreme Gradient Boosting (XGBoost), the Multi-Layer Perceptron (MLP), CNN and Support Vector Machine (SVM) algorithms. The database is managed by the authors to classify the applications. Experimental results show that the suggested model performance is better than the other algorithms.

In [14]. Authors study helps to detect DOS attacks and brute force attacks in Subtractive Fuzzy Cluster Means (SFCM) computer network systems by making FCM breaks down complex data into segments. The experiment demonstrates that the FCM (Fuzzy Cluster) achieved more efficient cluster results than the hard clusters (K-means) as well as the soft clusters (Hard and Soft). By combining the Subtractive Cluster for performances improvement with the FCM algorithm.

In [15]. The authors presented an algorithm to improve the efficiency and robustness of the Low-rate DoS (LDoS) detection system by combining PSD-enteropy and support vector machine (SVM). The proposed algorithm is fully utilized by both methods. Entropy application efficiently reduces the amount of calculations while SVM supports the proposed method by minimizing the dataset to the most pertinent characteristics. To measure entropy, two adaptive thresholds are set. The detection frequency and detection amount can be regulated by adjusting the two thresholds in the detection method. The thresholds have an 80% emphasis on reliability in order to achieve a higher detection rate. The use of PSD-entropy guarantees a time-complexity of the algorithm. Experience results show that 99,19% of LDoS attacks and O(n log n) time complexity in the best case can be detected in the approach proposed. This allows for a better detection of LDoS attacks and for potential LDoS attacks detection.

In [16]. The authors used data from SNMP-MIB to detect patterns of DOS attacks that may affect the network. To classify the dataset they used three machine learning algorithms (Random Forest, J48 Decision Tree and REP Tree). Two evaluators of attributes were used to exclude unnecessary variables and to obtain top 5 and top 3 variables, the two evaluators of attributes are InfoGain and ReliefF. The classifiers and attributes were applied to the IP group and the results showed that the application of the REP tree algorithm classifier provided the highest precision all the time in the top 5 and top 3 IP group.

In [17]. The authors implemented a network traffic detection system that distinguishes DOS attacks from normal traffic based on an adopted dataset. The results showed that in detecting ICMP Echo attack, HTTP Flood Attack, and Slowloris attack, the adopted algorithms with the ICMP variables achieved a high precision percentage with approximately 99.6%. In addition, the built prototype succeeded in varying normal traffic from different DOS attacks at a rate of 100%.

In [18]. The authors analysed latest tests of the identification system of DoS attacks. a smart detection / defense system is introduced for DoS assaults. The system took the DoS assault fitting and the usual traffic as feedback and then carried out the two-stage design. The detection model of the dos attacks based on the neural networks of BP was increased to 99.97% during the ŠIRT stage. In the second phase, a complex game theory-based defence strategy was used to raise the DoS attack detection rate to 99.98%. Eventually, the simulation results demonstrate rational system parameters like the evaluation coefficient and the JR score, and efficiency of the proposed scheme.

In [19]. The authors used Support Vector Machine (SVM) and C4.5 supervised learning algorithms on NSL KDD Dataset for successful DOS Attack classification.  A sniffer is also used to track IP packets in the network and detect malicious and natural traffic packets. The findings show the successful and accurate assessment results using the SVM and C 4.5 supervised learning algorithm.

In [20]. Thesis discusses how the computer network utilizes machine learning techniques to identify irregularities in real time. Although machine learning is not a new concept for this task, little literature is about it in real time. In order to demonstrate the usefulness of the algorithms provided, most machine learning research in anomaly detection of computer network is focused on KDD ' 99 data set. The reliance on this data set has created a scarcity of research papers on how network data can be obtained, features retrieved, or algorithms qualified for use in real-time networks. It was alleged that KDD ' 99 data set was not applied to real-time networks for discovery of anomalies. The study suggests how to use a dumb network to gather data and contrasts k-means cluster results, one group SVM and LSTM neural networking with recorded KDD ' 99 algorithm outcomes. The results demonstrate that KDD-based algorithms are less reliable but are related to the lack of sophistication in the information gathered.

In [21]. The authors proposed intrusion detection system for Denial of Service (DOS) and Distributed DOS (DDOS) attacks using Hellinger distance (HD). the Results indicate that the proposed mechanism performed reliable detection of DOS/DDOS flooding attacks.

In [22]. The authors introduced new concept that can support the right source of data while avoiding the issues associated with the binary decision, also aims to use the Fuzzy Rule Interpolation (FRI) with Simple Network Management Protocol (SNMP) and Management Information Base (MIB) parameters to implement a detection approach and identify anomalies. The intensity of the proposed approach for detection is based on adapting the parameters of SNMP-MIB to the FRI. the proposed method eliminates the time-consuming raw traffic processing component and requires comprehensive computational measures. It also eliminates the need for a definition of invasion based on a complete fuzzy rule. The suggested solution was validated and analysed using an SNMP-MIB open source dataset and the results shows that 93% detection rate was achieved.

In [23]. The authors used Artificial Neural Networks (ANN) machine learning to detect Denial of Service (DoS) and Distributed denial of Service (DDoS) attacks, the proposed method used the back propagation Bayesian Regularization (BR) and scaled back spreading algorithm (SCG). The results show that 99.6% of DoS / DDoD attacks using Bayesian regularization were successfully detected by the proposed method and 97.7% of the attacks in scaled concentration descent.

In [24]. The authors introduced an effective network attack detection system and types of attack identification using the Management Information Base (MIB) database associated with the Simple Network Management Protocol (SNMP) via machine learning techniques, and also investigates the effect of SNMP-MIB information on the detection of network anomalies. Three classifiers are used to construct the detection model, namely Random Forest, AdaboostM1 and MLP. Using various classifiers presents a comprehensive study on the effectiveness of data from SNMP-MIB in detecting different types of attack. The results show that the quality of each classifier varies with the Interface MIB group, where the accuracy level for the three classifiers ranged from high to low. With the Interface group 99.93 %, the Random Forest classifier achieved the highest precision level.

In [25]. The authors used Machine learning (ML) and Neural Network (NN) algorithms to detect DoS attacks.

The detection focused especially on DoS-Aggress Detection applications instead of on transport and DoS-Aggression network detection. In the experiment, the new DoS attack dataset is used. The trial divided the data set into separate fractions and for each algorithm the best fraction is found. RF and MLP. RF and MLP. The RF and MLP tests are compared and the results from RF are shown to be higher than MLP.

In [26]. The authors proposed multi-threaded Intrusion Detection System (IDS) framework that Pre-processes it, removes network functionality and transfers it to the intrusion detection classifiers. This uses both anomaly-based signature and complementary detection. Consequently, both known and unknown cloud attacks can be observed. Low-fake positive performance and low-fake negatives.

In [27]. The authors used machine learning classification techniques like AdaboostM1, Random Forest and MLP to build detection model for Denial of Service (DoS) flooding and brute force network attacks within MIB data related with (SNMP) protocol, they categorized the MIB variables to five groups (ICMP, Interface, IP, TCP and UDP) and showing how attacks effects on each party. The results show that the Interface and IP groups are the only groups among five MIB groups most affected by attacks of all types and less affected by ICMP, TCP and UDP, Random Forest classifier obtained the highest precision Rate with the IP group (100%) and with the Interface group (99.93%).

In [28]. The authors used machine learning algorithms (Neural Networks, K-Nearest Neighbours and C5.0) and Remote Monitoring (RMON) data polled from switch to detect DoS attacks that achieved low false alarm rate and high detection rate based on their traffic and laboratory environment, the results shows that Neural Networks achieved the highest accuracy comparing with the rest of algorithms used.

In [29]. The authors implemented various machine learning algorithms such as SVM, Logistic Regression, Decision-Tree Classification, K-Neighbour Classification, XGB Classification, and other algorithms for the prediction of flooded packet in SNMP networks to establish a model of maximum precise. the results show high accuracy on every algorithm used.

In [30]. The authors presented a new hybrid approach by tracking irregular network traffic to prevent a DDoS threat. The method reads traffic data and makes a system forecasts potential data and compares them with actual data to identify irregular traffic. this approach has never been used in this area before by combination from two methods (changing detection and traffic prediction) to detect any abnormal traffic in a network. The results show that this approach is 98.3% accurate and 100% sensitive. Also, the presented model captured all kinds of attacks with low false negative rates, each with its own behaviours.

In [31]. The authors used three machine learning algorithms like BayesNet, Support Vector Machine (SVM) and Multi-Layer Perceptron (MLP) to investigates the Intrusion Detection (ID) problem. The algorithms are used with a true, MIB data base that is obtained from the real world. A variety of solutions and technologies are used to improve accuracy in the detection process like Genetic Search (GS), Infogain (IG) and ReleifF (RF). The experiments show that the identification performance has improved with three practical choice methods. Genetic Search (GS), with bayesNet, MLP and SVM have a high precision frequency, especially the BayesNet (99,9%) with Genetic Search (GS).

In [32]. The authors proposed effective and scalable Network Intrusion Detection System based on deep learning approach. They used Self-Teaching Learning (STL), a methodology based on deep learning, on NSL-KDD-a network invasion benchmark dataset. The results show that STL achieved better recall values (95.95%) as compared to Soft-Max Regression (SMR) (63.73%)

In [33]. The authors used hybrid detection mechanism from Correlation Analysis and triangle-area to detect Denial of Service attack, the system was able to Differentiate known as well as unknown attacks from legitimate network traffic.

In [34]. The authors implemented a DoS/DDoS flood attack detection system based on SNMP MIB data which selects active MIB variables and compares different classification algorithms like Neural network, Bayesian network and C4.5 tree based on selected variables. The results show that Neural network have the highest accuracy with 99.035% detection rate.

In [35]. The authors proposed using MIB+, the standard MIB extension to construct a computer system and network intrusion detection profile. The proposed system of intrusion detection is based on network IDS, which only detects intrusions in the network. Using SNMP and MIB+ objects, it collects network traffic data and detects SYN floods, floods and Null scans. A comparison of the current traffic information from the profile is used for the process for evaluating an anomaly in the network. Then use a decision function to tell whether the results show an attack or not. Experiment results shows that MIB+ provides a high performance and precision in the proposed system.

In [36]. The authors proposed an algorithm of irregular traffic detection using the MIB artefacts in IP class. This algorithm allows a traffic collection cycle to be shortened without affecting the performance of traffic collection servers. Thus, a comprehensive and quick analysis online can be rendered and traffic can be analysed in depth by operators. The results show that with the exception of a couple of router models and switches, the algorithm can be used for most IP networking devices. The accuracy of the estimate is also well established in this algorithm and can be improved if the IP group traffic threshold are correctly set.

A significant finding in the SNMP-MIB-based attack identification literature is that most of the above experiments are restricted to certain forms and a small number of attacks. Furthermore, several of these experiments concentrated on identifying anomalous traffic as distinct from normal traffic without disregarding the type of attack. SNMP-MIB as a rich data source for increased security in computer networks still must be manipulated. Here we intend to offer a high detection and accuracy approach to detect attacks with the SNMP-MIB data compared to other approaches. Our work uses MIB data to identify several different attacks dependent on computer teaching techniques (HTTP flood, UDP flood, TCP-SYN, ICMP-ECHO, Slowpost, Brute Force, slowloris).

### III. PROPOSED MODEL

a) SNMP-MIB Dataset

In this experiment, the proposed framework is tested by a new SNMP-MIB dataset [6]. The reason of using this dataset in our experiment that it has unique records, simulates a realistic network traffic and contains about 34 MIB variable records. The assault details were obtained in 6 forms of DoS (TCPSYNs, UDPs, ICMP-ECHOs, HTTPs, Slowlooris, Slowpost), and Brute Force assaults. For the MIB dataset and the records cases as described in Table 1, the dataset contains 4998 MIB records, more info about dataset are possible in [6].

**Table 1.** The number of generated records by Traffic type [6].

|    | Type of traffic | Number of records |
|----|-----------------|-------------------|
| 1. | Normal          | 600               |
| 2. | TCP-SYN         | 960               |
| 3. | UDP flood       | 773               |
| 4. | ICMP-ECHO       | 632               |
| 5. | HTTP flood      | 573               |
| 6. | Slowloris       | 780               |
| 7. | Slowpost        | 480               |
| 8. | Brute Force     | 200               |

b) Machine Learning Classifiers

Classification is one of the most commonly used techniques for supervised data mining. The process is to find a model describing the data classes or concepts. The main objective is to simulate the entity type using the undefined class mark model. The classification model is created through trainings and the derived models, such as table, trees, or rules, can be presented in many forms. This paper concentrates on techniques such as DecisionTable, JRip, OneR, PART and ZeroR for the classification of rules.

1) DecisionTable

Decision tables are a descriptive visual representation of the behavior to be carried out depending on the conditions. These are algorithms whose output represents a set of actions. The information given in the decision tables could also be interpreted as decision trees or as a series of if-then-else and switch-case statements in a programming language For describing and analysing situations, table representation is used by decision tables (DTs). The decision i.e. action is taken based on the number and interrelationships of conditions [37].

2) JRIP

William W. Cohen's proposed JRIP. JRIP is a popular classificator algorithm, sometimes called as RIPPER. In JRIP instances, data sets are progressively analyzed, a set of rules are created for a certain hazard dataset. JRIP algorithm (RIPPER) manages increasing database dataset and outputs a set of rules with all the class characteristics. Then the next class is assessed and follows the same procedure as before, until all classes are protected [38].

3) OneR

OneR, referred to "One Rule" is the easiest method for discerning attributes but precise grading algorithm which generates one rule for each data predictor, then selects the rule with the smallest total error as its "one rule" In order to create a predictor rule, we create a frequency table against the target for every predictor. It was shown that OneR only slightly less accurate rules are produced than state-of - the-art classification algorithms while simple rules for humans are produced [39].

4) PART

PART is a standardized rule-making process combines a strategy of division and conquest. In each step, incomplete C45 tree is built, the leaves will be converted into rules after rule creation is created. PART embraces all class types such as Binary and Nominal classes, including all attributes [40].

5) ZeroR

This is the simplest way to just forecast the type of plurality (class), depends on a goal and avoids other predictors. It uses more information about a particular problem in order to create a rule to predict. Depending on the type of problem [39].

These classifiers were applied to our SNMP-MIB dataset to classify attacks and ordinary traffic and then measured for accuracy to detect various types of network attacks with machine learning technology in order to investigate the ability and efficiency of SNMP-MIB data.

IV. EXPERIMENTS

a) Implementation

For this study, our MIB data set was analysed in the identification and analysis of attacks with computer training techniques. We initially split the MIB dataset arbitrarily into a training dataset of 70 percent (3498 records) and a testing dataset of 30 percent (1500 records) for tests on classifiers. Both datasets contain normal and the other seven attack classes. In this paper, we applied DecisionTable, JRip, OneR, PART and ZeroR classifiers to the MIB dataset. The approach used the MIB system variables and then the group classifying algorithms. Our studies are carried out using a WEKA 3.8.3 toolkit [41] for open source data mining. Originally developed at the University of Waikato in New Zealand, WEKA (Waikato Environment for Knowledge Analysis). This application has been written in Java and has a series of algorithms for machine learning and data mining to pre-process, cluster, ranking and other results [41].

b) Performance Evaluation Metrics

In our work, we have used several important metrics to detect and classify the different types of attacks based on the MIB data set accurately. The classifiers performance like F-Measure, accuracy, precision and recall are calculated using a well-known evaluation criterion as shown from (1) to (4). Precision is the percentage of the correctly labelled positive results, and the Recall is the number of positive tests, which has been accurately identified as false. The true positive figure is related to, and the sensitive factor is also defined. Based on precision and retrieval methods, F-Measure is a measure of accuracy of classification type. It is a weighted average of the accuracy and recall measurements. In addition, confusion matrix (Table 2) is used to measure classifier efficiency and to display the classification process results; this can be seen using details on current and expected sample classifications used.

$$Precision = \frac{TP}{TP + FP} \quad (1)$$

$$Recall = \frac{TP}{TP + FN} \quad (2)$$

$$F\text{-}Measure = \frac{2 \cdot Precision \cdot Recall}{Precision + Recall} \quad (3)$$

$$Accuracy = \frac{TP + FN}{TP + FP + TN + FN} \quad (4)$$

**Table 2.** The confusion matrix for two classes

| Actual class | Predicted class | |
|---|---|---|
| | Positive | Negative |
| Positive | TP | FP |
| Negative | FN | TN |

Given that F-Measure is an evaluation measure, which depends on the precision and the recall metrics, hence it will be considered in all our experimental results as shown in the next section.

V. EXPERIMENTAL RESULTS AND DISCUSSION

The experimental results for the proposed method that uses a complete MIB dataset (same data as training and tests) are defined in this portion. The paper we use Interface group which contains the associated MIB variables, then used the DecisionTable, JRip, OneR, PART and ZeroR classification algorithms. The goal is to research and show the effect of the five groups on the detection of attacks using the machine learning classifiers.

a) Experimental Results with Interface MIB Group

In this experiment, the DecisionTable, JRip, OneR, PART and ZeroR classifiers are applied and measured on the Interface group dataset. Figures 2, 3, and 4 display the Precision, Recall, and F-measure performance of the classifiers based on MIB interface variables.

The results in Figure 2 shows that precision results are very high for DecisionTable, JRip, OneR and PART classifiers in all types of attack except for the ZeroR classifier that just achieved very low precision in TCP-SYN attack. Figure 3 indicate that DecisionTable, JRip, OneR and PART classifiers achieved very high results in correctly identifying the normal traffic records, as well as attacks records in the testing set, with high recall values from the results, JRip and PART classifiers succeeded in identifying most types of attacks, followed by DecisionTable, OneR classifiers, ZeroR classifier achieved very low performance in normal traffic and attacks detection with the interface group. More specifically, the JRip and PART classifiers achieved the best recall results for all types of attacks. Figure 3 shows F-measure results with Interface group.

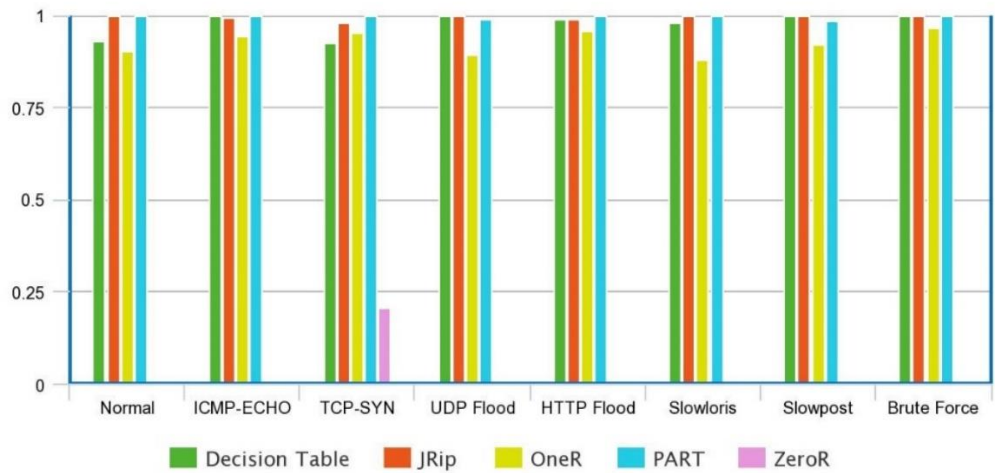

*Figure 2.* Precision results with Interface group.

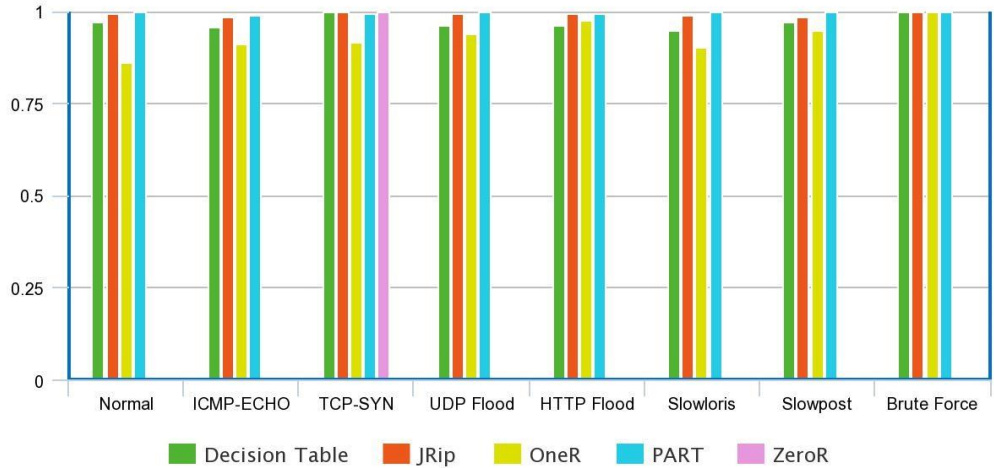

*Figure 3. Recall results with Interface group.*

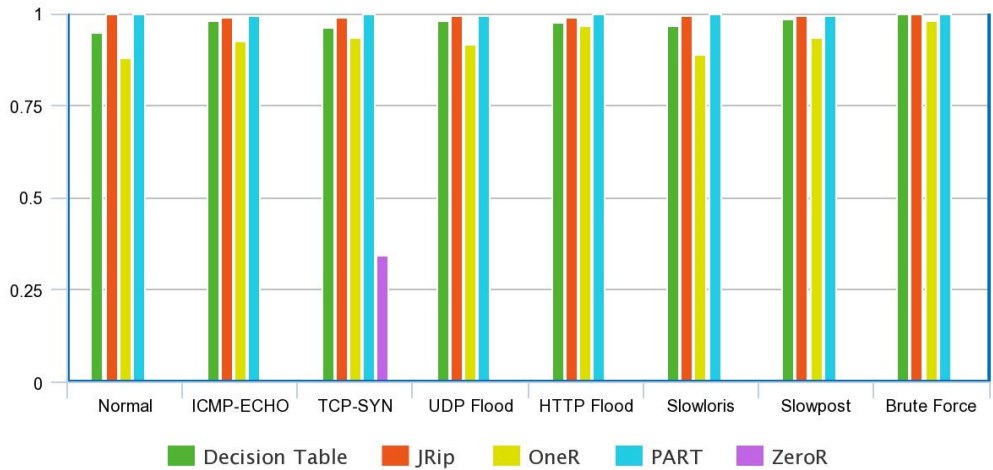

*Figure 4. F-measure results with Interface group.*

Figure 5 shows experimental results for the accuracy rate of all three Interface MIB Group Dataset Classifiers. As shown, the results indicate that the PART classification was outperformed with the Interface MIB group dataset by classifying intrusions (attacks), followed by the JRip classifier, which produced high accuracy results and very close to those obtained by PART. The ZeroR classifier, however, shows lower levels of accuracy in relation to the other classifiers with the interface MIB group dataset.

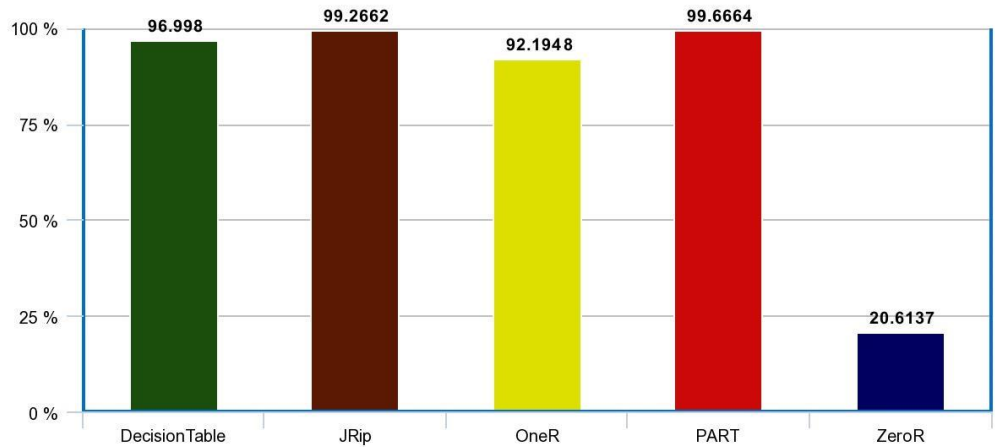

*Figure 5.* Accuracy rate for the five classifiers with interface group.

VI. CONCLUSION

In this paper, we introduced a network attack detection approach focused on SNMP-MIB knowledge through the implementation of Rule-based machine learning techniques. Our work aimed at demonstrating SNMP-MIB data capability and efficiency in the detection of network abnormalities by detecting the highest possible numbers of the most common and modern attacks that may take place on the interface layer. Five classification algorithms were included in our methodology, namely DecisionTable, JRip, OneR, PART and ZeroR. In our approach, we used the interface MIB variables, which included several associated MIB variables. Classification algorithms were then used to show the group how attacks affect it and therefore the Interface MIB Group's efficacy was determined in the detection of anomalies. From the results of this approach, we found that the performance of each classifier differs in the MIB interface group, where the exactness of the three classifiers varied from high to low. PART classifier achieved 99.7% accuracy which is highest value with the Interface group. We have also found from the results that the Interface MIB Group was affected by all forms of attack. The overall conclusion is that using SNMP-MIB data with machine learning techniques is a very effective way to detect anomalies in a network and affects on the security of it.